\documentstyle[pic]{article}

\begin{document}

\newcommand{\Pclass}{{\sc P}}
\newcommand{\NP}{{\sc NP}}
\newcommand{\BQP}{{\sc BQP}}
\newcommand{\QMA}{{\sc QMA}}

\title{Spin Systems and Computational Complexity}

\author{Daniel Gottesman}

\maketitle

What is the connection between a cathedral's stained glass window and the world's hardest Sudoku puzzle?
They are more closely connected than you might think.  Glass (not just stained glass) differs from most materials studied by physicists in that it has structure, but not a regular one.  The elemental composition of glass is not very different from quartz, but in quartz, the atoms are arranged in a regular crystalline structure, and rearranging the atoms to change the structure incurs a large energy cost.  In contrast, in glass, there are many different arrangements of atoms with about the same energy.  If liquid silica is cooled slowly, it can crystallize into quartz, but if it cools rapidly, the result is glass.  One atomic configuration is selected, more or less at random, when the glass cools, but it is not necessarily the lowest-energy one.


A similar phenomenon can occur with ``spin systems,'' systems where the atoms are all stationary and only the direction of their spins varies from location to location.  The direction of spin for a classical system is towards the north pole of its rotation axis, and a quantum spin similarly has a direction, though there is no easy interpretation as rotation.  A magnet is essentially a spin system: the atoms are arranged according to the structure of iron or whatever material composes the magnet.  However, each atom has a spin and an associated magnetic field.  The magnetic field from each atom interacts with the spin of nearby atoms.  In a ``ferromagnet,'' the spins and magnetic fields from different atoms tend to line up in the same direction.  They therefore reinforce each other, and add together to produce a field much larger than the magnetic field of any individual atom.  Iron is a ferromagnetic material, and permanent magnets result from an interaction of this type.  Some other materials are ``antiferromagnets.''  In an antiferromagnet, the magnetic fields of neighboring atoms tend to face opposite directions, and therefore cancel out.  Antiferromagnets are less common than ferromagnets, but they do occur naturally.  In a ferromagnet, the lowest-energy state (the ``ground state'') is just to line up all the spins to point the same way, whereas in an antiferromagnet, the ground state is for the spins to alternate which direction they face, forming a checkerboard pattern.

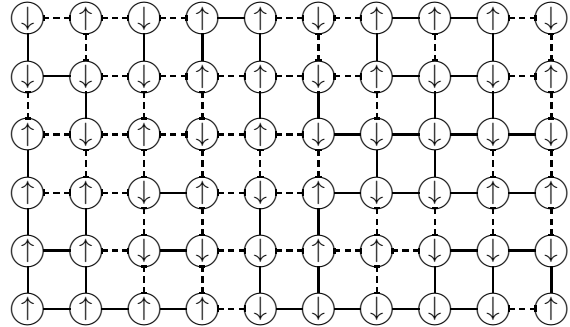
\begin{figure}

\begin{centering}

\begin{picture}(208,120)

\multiput(5,5)(22,0){10}{\circle{12}}
\multiput(5,27)(22,0){10}{\circle{12}}
\multiput(5,49)(22,0){10}{\circle{12}}
\multiput(5,71)(22,0){10}{\circle{12}}
\multiput(5,93)(22,0){10}{\circle{12}}
\multiput(5,115)(22,0){10}{\circle{12}}

\put(5,11){\line(0,1){10}}
\put(5,33){\line(0,1){10}}
\put(5,55){\line(0,1){10}}
\put(5,77){\dashbox{2}(0,10){ }}
\put(5,99){\line(0,1){10}}

\put(27,11){\line(0,1){10}}
\put(27,33){\line(0,1){10}}
\put(27,55){\dashbox{2}(0,10){ }}
\put(27,77){\line(0,1){10}}
\put(27,99){\dashbox{2}(0,10){ }}

\put(49,11){\dashbox{2}(0,10){ }}
\put(49,33){\dashbox{2}(0,10){ }}
\put(49,55){\dashbox{2}(0,10){ }}
\put(49,77){\dashbox{2}(0,10){ }}
\put(49,99){\line(0,1){10}}

\put(71,11){\dashbox{2}(0,10){ }}
\put(71,33){\dashbox{2}(0,10){ }}
\put(71,55){\dashbox{2}(0,10){ }}
\put(71,77){\dashbox{2}(0,10){ }}
\put(71,99){\line(0,1){10}}

\put(93,11){\line(0,1){10}}
\put(93,33){\line(0,1){10}}
\put(93,55){\dashbox{2}(0,10){ }}
\put(93,77){\line(0,1){10}}
\put(93,99){\line(0,1){10}}

\put(115,11){\line(0,1){10}}
\put(115,33){\line(0,1){10}}
\put(115,55){\dashbox{2}(0,10){ }}
\put(115,77){\line(0,1){10}}
\put(115,99){\dashbox{2}(0,10){ }}

\put(137,11){\dashbox{2}(0,10){ }}
\put(137,33){\dashbox{2}(0,10){ }}
\put(137,55){\line(0,1){10}}
\put(137,77){\dashbox{2}(0,10){ }}
\put(137,99){\line(0,1){10}}

\put(159,11){\line(0,1){10}}
\put(159,33){\dashbox{2}(0,10){ }}
\put(159,55){\line(0,1){10}}
\put(159,77){\line(0,1){10}}
\put(159,99){\dashbox{2}(0,10){ }}

\put(181,11){\line(0,1){10}}
\put(181,33){\dashbox{2}(0,10){ }}
\put(181,55){\dashbox{2}(0,10){ }}
\put(181,77){\line(0,1){10}}
\put(181,99){\line(0,1){10}}

\put(203,11){\line(0,1){10}}
\put(203,33){\dashbox{2}(0,10){ }}
\put(203,55){\dashbox{2}(0,10){ }}
\put(203,77){\dashbox{2}(0,10){ }}
\put(203,99){\dashbox{2}(0,10){ }}

\put(11,5){\line(1,0){10}}
\put(11,27){\line(1,0){10}}
\put(11,49){\dashbox{2}(10,0){ }}
\put(11,71){\dashbox{2}(10,0){ }}
\put(11,93){\line(1,0){10}}
\put(11,115){\dashbox{2}(10,0){ }}

\put(33,5){\line(1,0){10}}
\put(33,27){\dashbox{2}(10,0){ }}
\put(33,49){\dashbox{2}(10,0){ }}
\put(33,71){\dashbox{2}(10,0){ }}
\put(33,93){\dashbox{2}(10,0){ }}
\put(33,115){\dashbox{2}(10,0){ }}

\put(55,5){\line(1,0){10}}
\put(55,27){\line(1,0){10}}
\put(55,49){\line(1,0){10}}
\put(55,71){\dashbox{2}(10,0){ }}
\put(55,93){\dashbox{2}(10,0){ }}
\put(55,115){\dashbox{2}(10,0){ }}

\put(77,5){\dashbox{2}(10,0){ }}
\put(77,27){\dashbox{2}(10,0){ }}
\put(77,49){\dashbox{2}(10,0){ }}
\put(77,71){\dashbox{2}(10,0){ }}
\put(77,93){\dashbox{2}(10,0){ }}
\put(77,115){\line(1,0){10}}

\put(99,5){\line(1,0){10}}
\put(99,27){\dashbox{2}(10,0){ }}
\put(99,49){\dashbox{2}(10,0){ }}
\put(99,71){\dashbox{2}(10,0){ }}
\put(99,93){\dashbox{2}(10,0){ }}
\put(99,115){\dashbox{2}(10,0){ }}

\put(121,5){\line(1,0){10}}
\put(121,27){\dashbox{2}(10,0){ }}
\put(121,49){\line(1,0){10}}
\put(121,71){\line(1,0){10}}
\put(121,93){\dashbox{2}(10,0){ }}
\put(121,115){\dashbox{2}(10,0){ }}

\put(143,5){\line(1,0){10}}
\put(143,27){\dashbox{2}(10,0){ }}
\put(143,49){\line(1,0){10}}
\put(143,71){\line(1,0){10}}
\put(143,93){\line(1,0){10}}
\put(143,115){\line(1,0){10}}

\put(165,5){\line(1,0){10}}
\put(165,27){\line(1,0){10}}
\put(165,49){\line(1,0){10}}
\put(165,71){\line(1,0){10}}
\put(165,93){\line(1,0){10}}
\put(165,115){\line(1,0){10}}

\put(187,5){\dashbox{2}(10,0){ }}
\put(187,27){\line(1,0){10}}
\put(187,49){\line(1,0){10}}
\put(187,71){\line(1,0){10}}
\put(187,93){\dashbox{2}(10,0){ }}
\put(187,115){\dashbox{2}(10,0){ }}

\put(5,5){\makebox(0,0){$\uparrow$}}
\put(5,27){\makebox(0,0){$\uparrow$}}
\put(5,49){\makebox(0,0){$\uparrow$}}
\put(5,71){\makebox(0,0){$\uparrow$}}
\put(5,93){\makebox(0,0){$\downarrow$}}
\put(5,115){\makebox(0,0){$\downarrow$}}

\put(27,5){\makebox(0,0){$\uparrow$}}
\put(27,27){\makebox(0,0){$\uparrow$}}
\put(27,49){\makebox(0,0){$\uparrow$}}
\put(27,71){\makebox(0,0){$\downarrow$}}
\put(27,93){\makebox(0,0){$\downarrow$}}
\put(27,115){\makebox(0,0){$\uparrow$}}

\put(49,5){\makebox(0,0){$\uparrow$}}
\put(49,27){\makebox(0,0){$\downarrow$}}
\put(49,49){\makebox(0,0){$\downarrow$}}
\put(49,71){\makebox(0,0){$\uparrow$}}
\put(49,93){\makebox(0,0){$\downarrow$}}
\put(49,115){\makebox(0,0){$\downarrow$}}

\put(71,5){\makebox(0,0){$\uparrow$}}
\put(71,27){\makebox(0,0){$\downarrow$}}
\put(71,49){\makebox(0,0){$\uparrow$}}
\put(71,71){\makebox(0,0){$\downarrow$}}
\put(71,93){\makebox(0,0){$\uparrow$}}
\put(71,115){\makebox(0,0){$\uparrow$}}

\put(93,5){\makebox(0,0){$\downarrow$}}
\put(93,27){\makebox(0,0){$\downarrow$}}
\put(93,49){\makebox(0,0){$\downarrow$}}
\put(93,71){\makebox(0,0){$\uparrow$}}
\put(93,93){\makebox(0,0){$\uparrow$}}
\put(93,115){\makebox(0,0){$\uparrow$}}

\put(115,5){\makebox(0,0){$\downarrow$}}
\put(115,27){\makebox(0,0){$\uparrow$}}
\put(115,49){\makebox(0,0){$\uparrow$}}
\put(115,71){\makebox(0,0){$\downarrow$}}
\put(115,93){\makebox(0,0){$\downarrow$}}
\put(115,115){\makebox(0,0){$\downarrow$}}

\put(137,5){\makebox(0,0){$\downarrow$}}
\put(137,27){\makebox(0,0){$\uparrow$}}
\put(137,49){\makebox(0,0){$\downarrow$}}
\put(137,71){\makebox(0,0){$\downarrow$}}
\put(137,93){\makebox(0,0){$\uparrow$}}
\put(137,115){\makebox(0,0){$\uparrow$}}

\put(159,5){\makebox(0,0){$\downarrow$}}
\put(159,27){\makebox(0,0){$\downarrow$}}
\put(159,49){\makebox(0,0){$\downarrow$}}
\put(159,71){\makebox(0,0){$\downarrow$}}
\put(159,93){\makebox(0,0){$\downarrow$}}
\put(159,115){\makebox(0,0){$\uparrow$}}

\put(181,5){\makebox(0,0){$\downarrow$}}
\put(181,27){\makebox(0,0){$\downarrow$}}
\put(181,49){\makebox(0,0){$\uparrow$}}
\put(181,71){\makebox(0,0){$\downarrow$}}
\put(181,93){\makebox(0,0){$\downarrow$}}
\put(181,115){\makebox(0,0){$\uparrow$}}

\put(203,5){\makebox(0,0){$\uparrow$}}
\put(203,27){\makebox(0,0){$\downarrow$}}
\put(203,49){\makebox(0,0){$\uparrow$}}
\put(203,71){\makebox(0,0){$\downarrow$}}
\put(203,93){\makebox(0,0){$\uparrow$}}
\put(203,115){\makebox(0,0){$\downarrow$}}

\end{picture}
\caption{A two-dimensional spin system with random magnetic and ferromagnetic bonds.  The solid lines indicate a ferromagnetic bond and the dotted lines indicate an antiferromagnetic interaction.  Each spin can point up or down.  This particular configuration has $15$ bond conditions violated, but there are many other configurations with the same number of incorrect bonds.}
\label{fig:spinglass}

\end{centering}

\end{figure}

Spin systems can display an enormous range of possible behaviors.  For instance, if we have a randomly mixed material so that some neighboring pairs of atoms have ferromagnetic interactions whereas other pairs have antiferromagnetic interactions (as in Fig.~\ref{fig:spinglass}), the system no longer has a nice regular lowest-energy state.  Instead, there is a complicated morass of different states, all of which have very similar energies.  A spin system that behaves like this is called a ``spin glass.''  While it is a subject of dispute whether there is an actual connection between the physics of spin glasses and of window glass, spin glasses are nonetheless a fascinating subject in their own right.\cite{spinglass}

To study spin systems, physicists generally simplify them further.  We assume the atoms have truly fixed locations, perhaps on a square or cubic lattice, and only the spin state of an atom can change.  Frequently, we assume that only adjacent spins can interact, and that spins which are further apart have no direct effect on each other.  These idealized spin systems can be classical or quantum.  In a classical spin system, each spin considered by itself has a definite direction, and we can describe the system's configuration at any given time just by specifying the direction of each spin.  In a quantum spin system, the spins behave quantum mechanically, and thus can be in entangled states, where it is not possible to specify the state of just one spin, only of all the spins collectively.


A spin glass does not naturally find its own ground state, but one might imagine that with the aid of a powerful computer, we could still learn the lowest possible energy state of a spin glass.  Not so.
Computers have advanced dramatically in power in the past few decades, but there are still problems that we do not know how to solve.  Indeed, we believe that some problems, including finding the ground state energy of some spin glasses, are inherently hard to solve, and that future advances in computer engineering will still not let us solve the hardest examples.

Sudoku is another example of a computationally hard problem.  (For those who are unfamiliar with it, see figure~\ref{fig:sudoku} for an example.)  The examples of Sudoku presented as puzzles in newspapers and elsewhere are designed to be solvable, but if you remove that crutch and generalize to larger $k^2 \times k^2$ grids filled with numbers from $1$ to $k$, many examples of the puzzle will be too difficult to solve even with the aid of the world's largest computers.

\begin{figure}

\begin{centering}

\begin{picture}(155,155)

\multiput(10,10)(15,0){10}{\line(0,1){135}}
\multiput(10,10)(0,15){10}{\line(1,0){135}}

\small

\put(10,10){\makebox(15,15){3}}
\put(10,25){\makebox(15,15){6}}
\put(10,70){\makebox(15,15){8}}
\put(10,85){\makebox(15,15){2}}
\put(10,100){\makebox(15,15){1}}
\put(10,115){\makebox(15,15){7}}
\put(10,130){\makebox(15,15){4}}

\put(25,10){\makebox(15,15){1}}
\put(25,25){\makebox(15,15){4}}
\put(25,55){\makebox(15,15){7}}
\put(25,85){\makebox(15,15){9}}
\put(25,115){\makebox(15,15){8}}

\put(40,25){\makebox(15,15){2}}
\put(40,55){\makebox(15,15){1}}
\put(40,70){\makebox(15,15){4}}
\put(40,85){\makebox(15,15){6}}
\put(40,115){\makebox(15,15){9}}

\put(55,10){\makebox(15,15){8}}
\put(55,25){\makebox(15,15){1}}
\put(55,40){\makebox(15,15){3}}
\put(55,55){\makebox(15,15){4}}
\put(55,70){\makebox(15,15){2}}
\put(55,85){\makebox(15,15){7}}
\put(55,100){\makebox(15,15){5}}

\put(70,10){\makebox(15,15){9}}
\put(70,25){\makebox(15,15){7}}
\put(70,40){\makebox(15,15){2}}
\put(70,70){\makebox(15,15){1}}
\put(70,85){\makebox(15,15){5}}
\put(70,115){\makebox(15,15){3}}
\put(70,130){\makebox(15,15){8}}

\put(85,25){\makebox(15,15){5}}
\put(85,40){\makebox(15,15){6}}
\put(85,70){\makebox(15,15){9}}
\put(85,85){\makebox(15,15){8}}
\put(85,100){\makebox(15,15){7}}
\put(85,115){\makebox(15,15){2}}
\put(85,130){\makebox(15,15){1}}

\put(100,25){\makebox(15,15){9}}
\put(100,40){\makebox(15,15){4}}
\put(100,55){\makebox(15,15){2}}
\put(100,70){\makebox(15,15){7}}
\put(100,85){\makebox(15,15){1}}
\put(100,100){\makebox(15,15){8}}
\put(100,115){\makebox(15,15){5}}
\put(100,130){\makebox(15,15){3}}

\put(115,10){\makebox(15,15){2}}
\put(115,25){\makebox(15,15){8}}
\put(115,55){\makebox(15,15){9}}
\put(115,70){\makebox(15,15){5}}
\put(115,85){\makebox(15,15){3}}
\put(115,115){\makebox(15,15){4}}

\put(130,10){\makebox(15,15){5}}
\put(130,25){\makebox(15,15){3}}
\put(130,40){\makebox(15,15){7}}
\put(130,55){\makebox(15,15){8}}
\put(130,70){\makebox(15,15){6}}
\put(130,85){\makebox(15,15){4}}
\put(130,100){\makebox(15,15){9}}
\put(130,130){\makebox(15,15){2}}

\large

\put(10,40){\makebox(15,15){\bf 9}}
\put(10,55){\makebox(15,15){\bf 5}}

\put(25,40){\makebox(15,15){\bf 5}}
\put(25,70){\makebox(15,15){\bf 3}}
\put(25,100){\makebox(15,15){\bf 2}}
\put(25,130){\makebox(15,15){\bf 6}}

\put(40,10){\makebox(15,15){\bf 7}}
\put(40,40){\makebox(15,15){\bf 8}}
\put(40,100){\makebox(15,15){\bf 3}}
\put(40,130){\makebox(15,15){\bf 5}}

\put(55,115){\makebox(15,15){\bf 6}}
\put(55,130){\makebox(15,15){\bf 9}}

\put(70,55){\makebox(15,15){\bf 6}}
\put(70,100){\makebox(15,15){\bf 4}}

\put(85,10){\makebox(15,15){\bf 4}}
\put(85,55){\makebox(15,15){\bf 3}}

\put(100,10){\makebox(15,15){\bf 6}}

\put(115,40){\makebox(15,15){\bf 1}}
\put(115,100){\makebox(15,15){\bf 6}}
\put(115,130){\makebox(15,15){\bf 7}}

\put(130,115){\makebox(15,15){\bf 1}}

\linethickness{2pt}
\multiput(10,10)(45,0){4}{\line(0,1){135}}
\multiput(10,10)(0,45){4}{\line(1,0){135}}

\end{picture}
\caption{An example solved Sudoku problem.  The values and locations of the large bold numbers are the input to the problem.  The goal is to fill in the remaining locations so that each row, column, and $3 \times 3$ subgrid must contain exactly one of each digit from $1$ to $9$.  The values and locations of the small numbers provide a witness: With them in place, it is easy to check that this is a valid solution.  Given only the bold large numbers, however, it is difficult to find a solution.}
\label{fig:sudoku}

\end{centering}

\end{figure}

Computer scientists formalize the relative difficulty of various computational problems by categorizing them into ``complexity classes.''  To determine what complexity class a problem belongs to, one needs to look at its behavior for very big examples of the problem.  Given any single input for the problem, there is just one answer --- one output --- and the amount of time to get that answer might depend on what information you start with and exactly how your computer works.  However, when you look at larger and larger inputs for a problem, finding the answer typically gets harder and harder, and the approximate {\em rate} at which it gets harder does not depend on these details.  For instance, \Pclass\ is the class of problems that are solvable in a time which is any polynomial in the size $n$ of the input, be it $n^2$ or $n^{200}$.  The exact polynomial rate might depend on how your computer is built, but the fact that the growth is polynomial in most cases does not.%
\footnote{There is one major exception: a {\em quantum computer} can solve some problems in polynomial time which we believe cannot be solved in polynomial time on a regular classical computer.  \Pclass\ is the class of problems solvable in polynomial time on a classical computer.}
A problem in \Pclass\  is generally considered to be solvable in a reasonable time, and problems which are not in \Pclass\ are considered to be hard.  Of course, this is just a simplification --- a time scaling of $n^{200}$ is enough to make the problem hard in practice, whereas a time of $\exp (n / 10^{200})$ will be over before you know it unless $n$ itself is ridiculously large.  In addition, the scaling refers to the difficulty of solving the very hardest inputs; for many, or even most, inputs, the difficulty may be much less.  Still, it seems to be a reasonable criterion, in that it is both well-defined (because \Pclass\ doesn't depend on exactly how you define ``computer''), and for most problems outside \Pclass, there seem to be some reasonable-size inputs for which we cannot solve the problem.

Another important complexity class is \NP, which, roughly speaking, is the class of problems that can be checked in a reasonable amount of time.  Sudoku is an example, along with many other interesting problems.  More specifically, \NP\ is composed of ``yes'' or ``no'' questions (e.g., does this Sudoku have a solution?).  If the answer for a specific input is ``yes,'' there must be some information, called a ``witness,'' that will enable you to check in polynomial time that the answer is indeed ``yes.''  If the answer is ``no,'' then no purported witness should pass this checking procedure.  For Sudoku, the witness is simply the solution. {\em Finding} the solution is hard, but if you are {\em given} the solution, you can easily check that it is valid.  Indeed, Sudoku is an example of an ``\NP-complete'' problem.\cite{sudoku}  \NP-complete problems are the hardest problems in \NP\ --- if you can efficiently solve an \NP-complete problem for all inputs, you can efficiently solve {\em any} problem in \NP.  It is a famous open question whether $\Pclass = \NP$.  We believe it does not, and that therefore the \NP-complete problems are hard.


One strategy you might adopt to solve an \NP-complete problem is to try different potential witnesses.  If you happen upon a correct witness, it is easy to check, and therefore you know the answer is ``yes.''  If you are unable to find a valid witness, you might conclude the answer is ''no.'' Of course, the number of potential witnesses is huge; there are exponentially many in the input size.  You might repeatedly modify a potential witness slightly, attempting to overcome its defects.  For instance, in Sudoku, you might take a solution which has a column with two $9$s and no $3$ and change one $9$ to a $3$.  This might create new errors, requiring further changes, but perhaps after a few changes there will be fewer errors than in your original failed solution.  Some version of this strategy works quite well for many particular inputs.  However, the strategy fails on the very hardest inputs, because there are very many nearly-correct witnesses, and it is very difficult to find the one true witness among the forest of false witnesses.  In the case of Sudoku, there could be many arrangements that are incorrect in just a few locations, but the true solution might be very different from the almost-correct ones.

This is precisely the phenomenon that prevents spin glasses from settling down to a single state: There are many low-energy states, but only one of those (or a few at most) has absolutely the smallest energy.  Indeed, the problem of finding the ground state of a classical spin glass is often an \NP-complete problem.  For instance, it is an \NP-complete problem to find the ground state of a spin system in three dimensions with some mix of ferromagnetic, antiferromagnetic, and zero interactions.\cite{barahona}


Quantum mechanics adds an additional twist.  A ``quantum computer'' is a computer whose memory and computational registers may contain quantum superpositions.  By taking advantage of this capability, a quantum computer can solve some problems which seem to be too hard for classical computers.  The complexity class \BQP\ is defined as the class of problems which can be solved in polynomial time on a quantum computer, and we believe \BQP\ is bigger than \Pclass.
%
%
For instance, we believe factoring is in \BQP\ but not in \Pclass:  Multiplying two large prime numbers together is easy (in \Pclass), but going the other way, finding the prime factors of a large number, is believed to be hard for a classical computer.  In contrast, a quantum computer could factor numbers in polynomial time.\cite{Shor}  Small quantum computers have been built, but it will still be decades before we can build one large enough to factor numbers that can't be factored with today's classical computers.

Even without quantum computers, we can study the new quantum complexity classes they suggest, such as \BQP, and try to apply any new insights we gain to better understand quantum physics.  There is also a quantum analogue of \NP\ called \QMA.
%
%
\QMA\ is the class of problems that can be efficiently checked on a quantum computer.  Just as we believe that \BQP\ is bigger than \Pclass, we believe that \QMA\ is bigger than \NP.  In other words, \QMA-complete problems are likely too hard to even be efficiently {\em checked} on a classical computer.  They are probably also too hard to solve efficiently with a quantum computer.  (Indeed, we believe quantum computers can't solve every problem in \NP\ either.)

Finding the ground state energy of a quantum spin glass is \QMA-complete. That is, if we could solve this problem, we could solve any problem in \QMA.  Thus, quantum spin glasses are even more difficult than their classical counterparts, which are only \NP-complete.  In addition, more quantum systems are hard than classical systems.  Finding the ground state energy of a $1$-D classical spin system is in \Pclass\ --- easy --- but finding the ground state energy of a $1$-D quantum spin system is \QMA-complete.\cite{QMA1D}

Now you know how a cathedral's stained glass window and the world's hardest Sudoku problem are related.  Glass is disordered because it has a multitude of nearly-optimal configurations, the same effect that makes Sudoku and some other computational problems intractable.  Of course, a stained glass window and a Sudoku puzzle are not identical: The difference is that the stained glass window is supposed to help you pray, whereas with a really hard Sudoku, you can only pray for help.

\section*{Acknowledgements}

The author thanks Lucy Zhang for helpful comments.  He is supported by CIFAR, NSERC, the Government of Canada through Industry Canada and the Province of Ontario through MRI.  This paper was written in part while the author was visiting KITP, which is supported by the NSF under Grant No.~NSF PHY05-51164.

\end{document}